# Light Curve Analysis of GSC 2750-0054 and GSC 03208-02644

M. M. Elkhateeb[1,2] and M. I. Nouh[1,2]

E-mail: abdo_nouh@hotmail.com

[1]Physics Department, College of Science, Northern Border University, Arar, Saudi Arabia

[2]Astronomy Department, National Research Institute of Astronomy and Geophysics, 11421, Helwan, Cairo, Egypt

**Abstract:**

We present the first photometric analysis for the newly discovered eclipsing binary systems of Algole-type GSC 2750-0054 and GSC 03208-02644. Our analysis was carried out by means of the most recent version of Wilson Devinney (WD) code, which applies the model atmosphere by Kurucz (1993) with prescription in pass band for the radiative treatment. The accepted light curve solutions reveal absolute physical parameters and the spectral classifications for the components are adopted. Distance to each system was calculated based on the parameters of the accepted photometric solutions. Comparison with the evolutionary models are presented.

## 1. Introduction

Modelling the light of the eclipsing binaries can provide precise results of the physical parameters and thus evaluate their evolutionary state. This is because the observed quantities like brightness, colour and radial velocity give strong constraints on the geometric configuration of a given system. One of the characteristic of the components in the detached and semi -detached binaries have not filled their Roche lobes yet and hence one could treat their evolution as single stars.

The binaries GSC 2750-0054 and GSC 03208-02644 are newly discovered short period eclipsing systems of Algole type (EA). The present paper is a continuation of the series concerning the photometric analysis of the newly discovered eclipsing binaries, Elkhateeb et al. (2015), Elkhateeb et al. (2015) Elkhateeb and Nouh (2015) and Elkhateeb et al.(2014)

The present paper consists of five sections as follows: Section 2 is devoted to the new times of minima. Section 3 deals with light curve modelling and their parameters. In section 4 we investigate the evolutionary status for both systems. Conclusions of the results are outlined in section 5.

## 2. Photometric data and minima times

### 2.1. GSC 2750-054

The system GS 2750-0054 (RA (2000), $23^h\ 00^{'}\ 07^{''}$, Dec (2000), $30^d\ 39^{'}\ 18^{''}$) was discovered as a variable star in 2013 by Nelson and Buchhein (2014), it was classified as detached system (EA) with period of $0.^d47187$. Observations by Nelson was carried out in the period from 3 to 26 August, 2013 using 33 cm f/4 Newtonian telescope on Paramount ME mount of Sylvester Robotic Observatory, with SBIG ST-10 XME CCD in clear filter only. The star TYC 2750 1818 was used as a comparison star. Buchhein observed the system in the period from 19 to 28 August 2013 in B, V, and R pass bands by means of 11-inch (28 cm) Schmidt-Cass at F/6.3 telescope of Altimira Observatory (Minor Planet Center Observatory Code G76) with ST-8XE CCD (plate scale = 1.1 arc-sec pixel).

A complete light curves were obtained in V and R pass bands. A total of six minima were estimated from Nelson and Buchhein observations using a Minima V2.3 package (Nelson 2006) which based on the Kwee and Van Worden (1956) fitting method. Table 1 represents the new minima together with the (O-C)'s and Epoch (E) calculated using the first ephemeris by Nelson (2014):

$$\text{Min I} = 2456630.7385 + 0.^d471877 \times E \qquad (1)$$

The estimated (O-C)'s are not sufficient to give any informations about the period behaviour of the system GSC 2750-0054.

### 2.2 GSC 03208-02644

The system GSC 328-02644 (RA (2000), $22^h\ 23^{'}\ 43^{''}$, Dec (2000), $41^d\ 19^{'}\ 07^{''}$), was detected and discovered as a variable star of detached type (EA) with period of $1.^d1579$ in 2010 by Liakos & Niarchos (2011) in the frames of V407 Lac. The star TYC 3208 2737-1 was used as a comparison star. They carried out the first observations for the system during the period from 21 July to 1 August 2010 in B and I (Bessell) pass bands using 0.2 m reflector f/5 telescope of the university of Athens observatory equipped with SBIG ST-8XMEi CCD camera.

A total of six new minima (two primaries and four secondaries) and the corresponding residuals (O-C)'s are calculated using the first linear ephemeris by Liakos and Niarchos (2011) given by:

$$\text{Min I} = 2455399.4818 + 1.^{d}1597 \times E \qquad (2)$$

Table 1. Times of minima for the systems GSC 2750 0054 and GSC 03208 02644.

| Star | HJD | Error | Filter | Type | E | (O-C) | Reference |
|---|---|---|---|---|---|---|---|
| GSC 2750 0054 | 2456507.8520 | ±0.0004 | Clear filter | I | -260 | 0.0375 | This paper |
| | 2456529.7942 | ±0.0002 | V | I | -214 | 0.0374 | This paper |
| | 2456529.7947 | ±0.0005 | R | I | -214 | 0.0379 | This paper |
| | 2456530.7375 | ±0.0004 | Clear filter | | -212 | 0.0369 | |
| | 2456532.8626 | ±0.0005 | R | II | -207.5 | 0.0386 | This paper |
| | 2456532.8629 | ±0.0005 | V | II | -207.5 | 0.0389 | This paper |
| GSC 03208 02644 | 2455399.4808 | ±0.0004 | B | I | 0.00 | 0.0000 | This paper |
| | 2455399.4820 | ±0.0004 | I | I | 0.00 | 0.0000 | This paper |
| | 2455402.3824 | ±0.0007 | B | II | 2.5 | 0.0014 | This paper |
| | 2455402.3809 | ±0.0004 | I | II | 2.5 | -0.0002 | This paper |
| | 2455410.4989 | ±0.0006 | B | II | 9.5 | -0.0001 | This paper |
| | 2455410.4979 | ± 0.0004 | I | II | 9.5 | -0.0011 | This paper |

## 3. Light Curve Analysis

The analysis of the observed Light curves of the systems GSC2750-0054 and GSC 03208-02644 were performed using 2009 version of Willson and Devinney code (windows interface version by B.Nelson (http://members.shaw.ca/bob.nelson/software1.htm). The initial value for the temperature of the primary component ($T_1$) is estimated using the (J-H) infrared colour index for each system listed in SIMBAD (http://simbad.u-strasbg.fr/simbad/sim-fbasic).

We used (J-H) colour index temperature relation by Tokunaga (2000) to estimate the corresponding temperature for each colour index. We analysed all individual observations of the observed light curves in each band. Bolometric albedo and gravity darkening were assumed for convective envelopes ($T_{eff} < 7500$). Hence we adopted $g_1 = g_2 = 0.32$ (Lucy 1967) and $A_1$

= $A_2$ = 0.5 (Rucinski 1969). Bolometric limb darkening values are adopted using tables of Van Hamme (1993) based on the logarithmic law for extinction coefficients. Through the light curve solution, the commonly adjustable parameters employed are, the orbital inclination (**i**), mass ratio (**q**), temperature of the secondary component (**$T_2$**), surface potential **$\Omega_1$** (only for the system GSC 03208-02644), **$\Omega_2$** (for both systems) and the monochromatic luminosity of the primary star (**$L_1$**). The relative brightness of the secondary component is calculated from the stellar models.

### 3.1. GSC 2750-0054

The light curve analysis for the system GSC 2750-0054 is performed using Buchhein (2014) observations in V and R passbands using the W-D code (Nelson 2009) through Mode 4 (semidetached). A set of parameters represent the observed light curves are estimated after some trails and lead to the best photometric fit listed in Table 2.

Table 2. Photometric solution for the systems GSC 2750-0054 and GSC 03208 02644.

| Parameter | GSC 2750 0054 | GSC 03208 02644 |
|---|---|---|
| $i$ ($^0$) | 82.54±0.33 | 86.91±0.16 |
| $g_1 = g_2$ | 0.32 | 0.32 |
| $A_1 = A_2$ | 0.5 | 0.5 |
| q = $M_2$ / $M_1$ | 0.6493±0.0047 | 0.6274±0.0007 |
| $\Omega_1$ | 3.1530 | 5.8399±0.0235 |
| $\Omega_2$ | 3.8111±0.0078 | 4.0880±0.0141 |
| $\Omega_{in}$ | 3.1530 | 3.1135 |
| $\Omega_{out}$ | 2.7767 | 2.7483 |
| $T_1$ (°K) | 4650* | 8170* |
| $T_2$ (°K) | 4152±10 | 7663±22 |
| $r_1$ pole | 0.3924±0.0023 | 0.1914±0.0022 |
| $r_1$ side | 0.4147±0.0026 | 0.1926±0.0022 |
| $r_1$ back | 0.4440±0.0025 | 0.1937±0.0023 |
| $r_2$ pole | 0.2437±0.0068 | 0.2146±0.0036 |
| $r_2$ side | 0.2484±0.0074 | 0.2174±0.0039 |
| $r_2$ back | 0.2568±0.0085 | 0.2224±0.0043 |
| $\sum (O-C)^2$ | 0.0202 | 0.2592 |

*Not adjusted

According to the accepted model, the primary component is more massive and hotter than the secondary one with temperature difference of about 500 $^0$K. Fig. 1 displays the observed light curves together with the synthetic curves in V and R passbands. Based on the estimated parameters, a three dimensional geometrical structure for the system GSC 2750-0054 are obtained by means of the software package Binary Maker 3.03 (Braad street and Steelman, 2004), and displayed in Fig. 2.

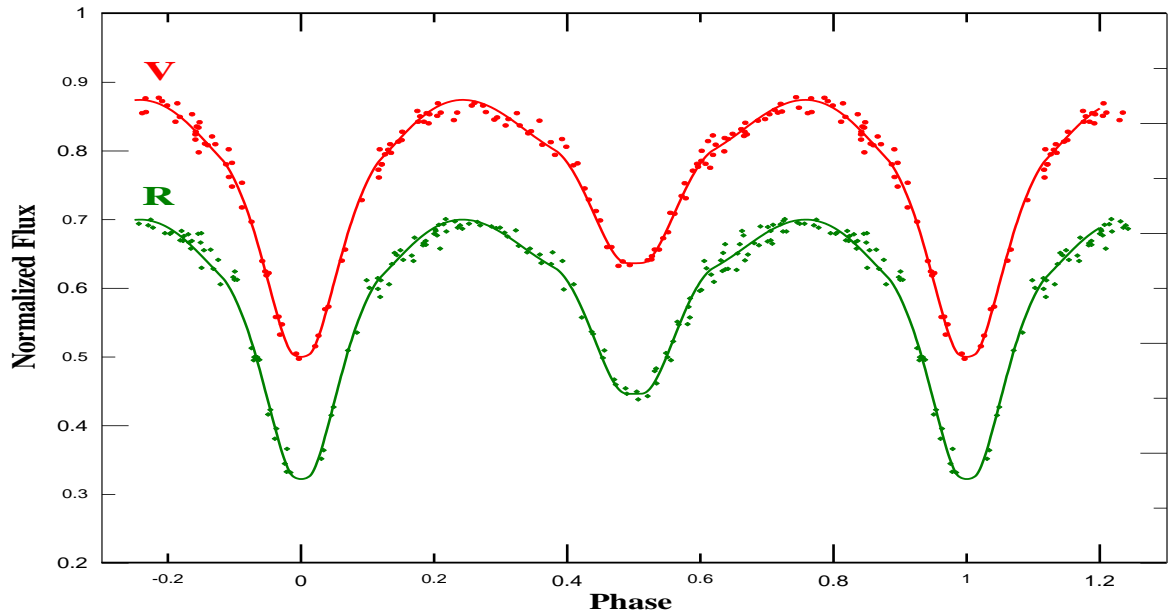

Fig. 1. Observed and synthetic light curves for the system GSC 2750 0054.

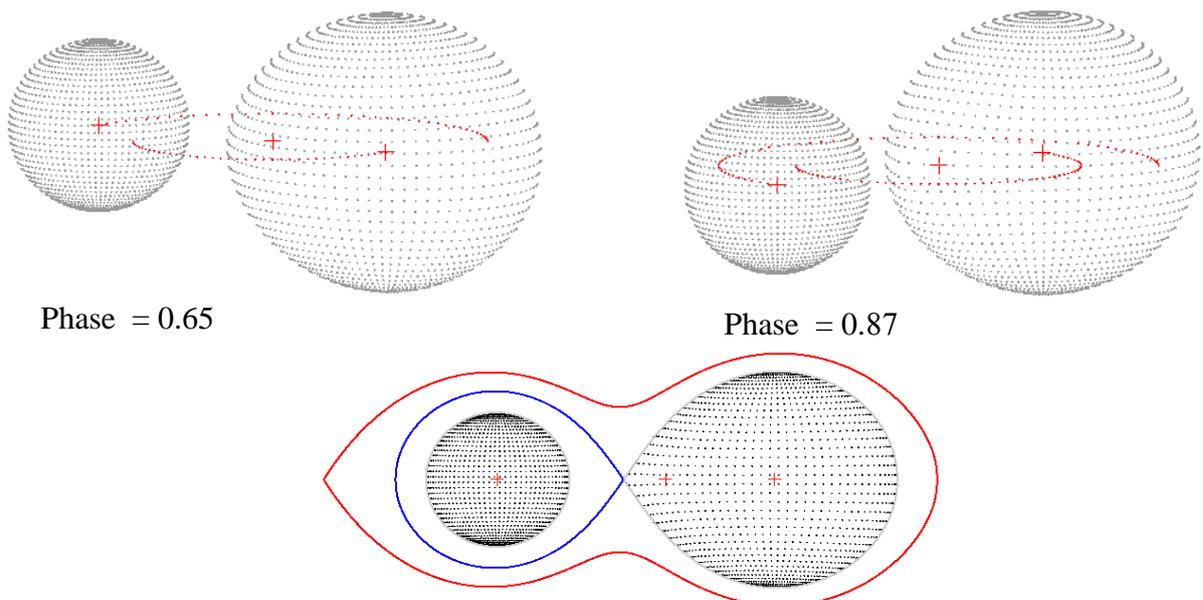

Fig. 2. Geometric structure of the binary system GSC 2750 0054.

## 3.2. GSC 03208-02644

We carried out the photometric solution to the system GSC 03208-02644 in B and I pass bands using Mode 2 (detached) of W-D code (Nelson, 2009). The parameters of the accepted model are listed in Table 2. The best light curve solution reveals that the primary component is hotter than the secondary one by about 500 $^0$K.

The reflected observed points in B and I pass bands are displayed in Fig. 5 together with the corresponding theoretical light curves. The corresponding three dimensional structure of the system GSC 03208-02644 (according to the parameters of the accepted model) is displayed in Fig. 6 with two hot spots on both components.

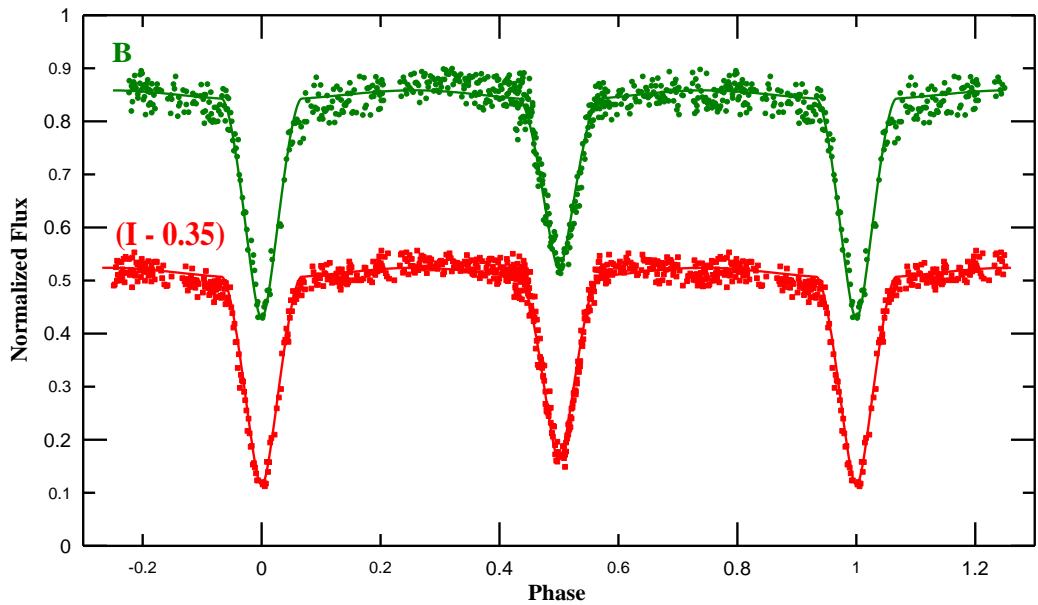

Fig. 3. Observed and synthetic light curves for the system GSC 03208 02644.

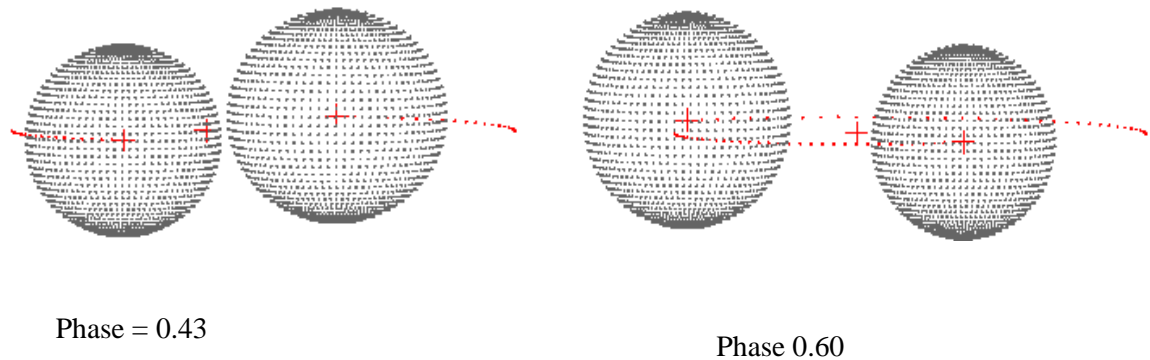

Phase = 0.43         Phase 0.60

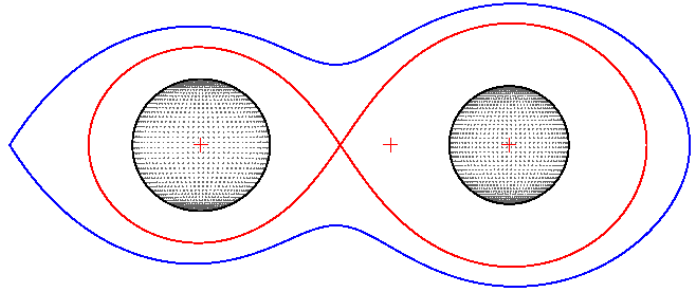

Fig. 4. Geometric structure of the binary system GSC 03208 02644.

## 4. Results and Conclusion

To determine the physical parameters of the two systems and due to the no-availability of radial velocity curves, we used the empirical $T_{eff}$ – mass relation by Harmanec, 1988. The estimated physical parameters reveals that the primary components in both systems are more massive than the secondary ones. Distances (*d*) to both systems were calculated by means of the estimated photometric and absolute properties ($d = 10^{(m-Mv + 5)/5}$), where *m* and $M_v$ are the apparent and absolute magnitudes respectively.

The calculations give the spectral types of the system GSC 2750-0054 as K3 and K7 and the distance as 116 ±5 pc. The physical parameters of the system are $M_1(M_\odot)$ =0.72±0.03, $M_2 (M_\odot)$= 0.47±0.02, $R_1(R_\odot)$ =0.81±0.03, $R_2(R_\odot)_2$= 0.62±0.03, $T_1(T_\odot)$=0.80±0.03, $T2(T_\odot)$=0.72±0.03, $L_1(L_\odot)$=0.27±0.01, $L_2(L_\odot)$=0.1±0.004, $M_{bol\_1}$=6.15±0.25 and $M_{bol\_2}$=7.22±0.29. The spectral types of the components of the system GSC 03208 02644 are A5 and A7, the distance to the system is 684±28 pc. The physical parameters are $M_1(M_\odot)$ =1.86±0.076, $M_2 (M_\odot)$= 1.21±0.049, $R_1(R_\odot)$ =1.84±0.075, $R_2(R_\odot)$= 1.75±0.072, $T_1(T_\odot)$=1.41±0.06, $T2(T_\odot)$=1.34±0.05, $L_1(L_\odot)$=13.58±0.55, $L_2(L_\odot)$=9.94±0.41, $M_{bol\_1}$=1.92±0.08 and $M_{bol\_2}$=2.26±0.092.

Based on the accepted photometric models, spectral classification for the components of the two systems are estimated and listed in Table 3. The physical parameters are used to investigate the evolutionary state of the systems. For this purpose we used the evolutionary tracks computed by Girardi et al. (2000) for both zero age main sequence stars (ZAMS) and thermal age main sequence stars (TAMS) with metalicity z = 0.019. The locations of the two systems on the M-R and M-L diagrams are shown in Figure 5 and Figure 6.

The primary component of the system GSC 03208-02644 lies on the ZAMS track for the M-R and M-L diagrams while the secondary component located near the TAMS track for both diagrams indicate that, the secondary is an evolved star.

For the binary system GSC 2750-0054, the two components are located near or on the TAMS tracks of the M-R and M-L diagrams which means that the two components are an evolve stars.

Also, we compared our results with the empirical mass–effective temperature relation for low and intermediate-mass stars based on data on detached double-lined eclipsing binaries.

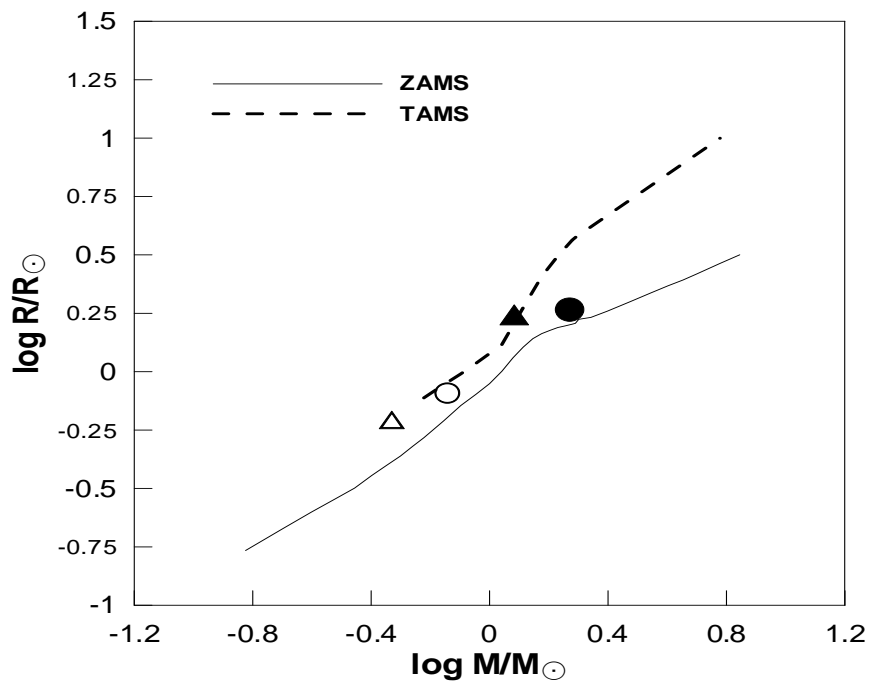

Figure 5: Positions of the two systems GSC 2750-0054 and GSC 03208 02644 on the mass–radius diagram of Girardi et al. (2000). Open symbols for GSC 2750-0054 and closed symbols for GSC 03208- 02644. Circles for the primary and triangles for the secondary.

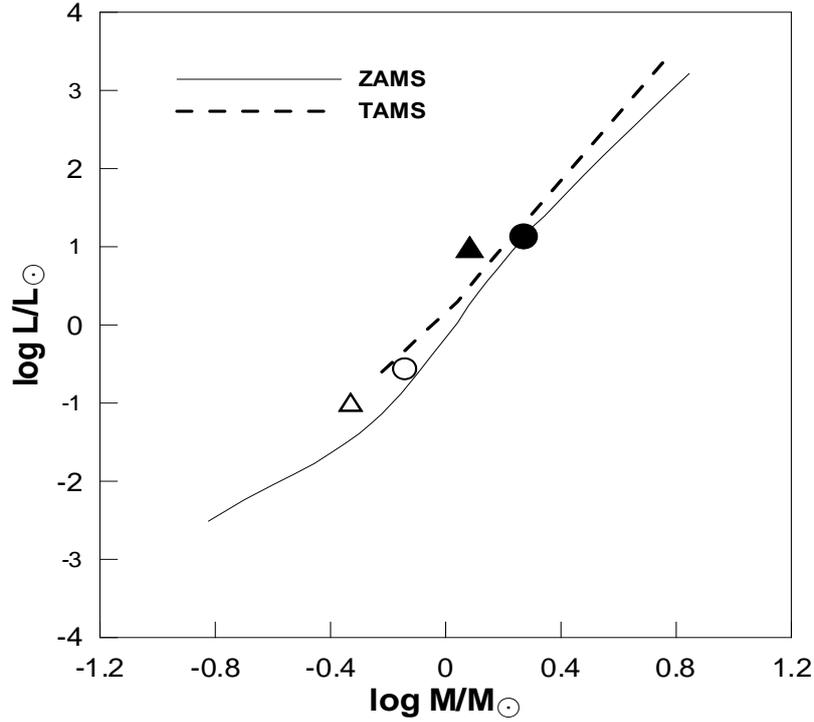

Figure 6: Positions of the two systems GSC 2750-0054 and GSC 03208 02644 on the mass–luminosity diagram of Girardi et al. (2000). Open symbols for GSC 2750-0054 and closed symbols for GSC 03208- 02644. Circles for the primary and triangles for the secondary.

In concluding the paper, we have performed light curve analysis for the newly discovered eclipsing binary systems GSC 2750-0054 and GSC 03208-02644 discovered in 2013 and 2010 respectively. A complete light curves are obtained for both systems and new times of minima are calculated. First photometric analysis for both systems reveals that the primary components in both systems are hotter and more massive than the secondary ones. Spectral classifications was adopted for each components based on the absolute parameters resulted from the accepted photometric solution. Distance to each system was calculated and a three dimensional structure is displayed.

As the two systems are detached and semi-detached, we expect that they could be modelled by the evolutionary models of single stars. Locations of the individual components on M-R, M-L and M-$T_{eff}$ diagrams give a preliminary result and need to be confirmed by more photometric and spectroscopic observations.

**Acknowledgments**

We wish to acknowledge the financial support of this work through Northern Border University, deanship of scientific research and higher education grant number 5-37-1436-5.